\documentclass[aps,prx,floatfix,letterpaper,
superscriptaddress,twocolumn,
  showpacs
longbibliography,nofootinbib,
]{revtex4}
  \usepackage[latin9]{inputenc}
\usepackage{amsmath,amsfonts,amsthm,amssymb}
\usepackage{color}
\usepackage[latin9]{inputenc}
\usepackage{graphicx}
\usepackage{tikz}
\usepackage{verbatim}
\usepackage{bm,latexsym,galois,euscript,dsfont}
\usepackage[all,cmtip,knot]{xy}
\usepackage{mathrsfs}
\usepackage[unicode=true,bookmarks=true,bookmarksnumbered=false,bookmarksopen=false,breaklinks=false,pdfborder={0 0 1},backref=false,colorlinks=true]{hyperref}

\hypersetup{
    colorlinks,
    linkcolor={red!60!black},
    citecolor={blue!90!black},
    urlcolor={blue!90!black}
}

\makeatletter
\@ifundefined{textcolor}{}
{%
 \definecolor{BLACK}{gray}{0}
 \definecolor{WHITE}{gray}{1}
 \definecolor{RED}{rgb}{1,0,0}
 \definecolor{GREEN}{rgb}{0,1,0}
 \definecolor{BLUE}{rgb}{0,0,1}
 \definecolor{CYAN}{cmyk}{1,0,0,0}
 \definecolor{MAGENTA}{cmyk}{0,1,0,0}
 \definecolor{YELLOW}{cmyk}{0,0,1,0}
}

\usepackage{amsfonts}\usepackage{tabularx}\usepackage{dcolumn}\usepackage{bm}\usepackage{graphicx}\usepackage{epstopdf}

\hypersetup{urlcolor=blue}

\theoremstyle{definition}

\newtheorem*{defn*}{Definition}
\newtheorem{thm}{Theorem}
\theoremstyle{remark}
\newtheorem*{pf}{Proof}
\newtheorem{example}{Example}

\makeatother

\begin{document}

\title{Entanglement Area Law for Shallow and Deep Quantum Neural Network States}
\date{\today}
\author{Zhih-Ahn Jia}
\email{\tt giannjia@foxmail.com,zajia@math.ucsb.edu}
\affiliation{CAS Key Laboratory of Quantum Information, School of Physics, University of Science and Technology of China, Hefei, Anhui, 230026, P.R. China}
\affiliation{CAS Center For Excellence in Quantum Information and Quantum Physics, University of Science and Technology of China, Hefei, Anhui, 230026, P.R. China}
\affiliation{Microsoft Station Q and Department of Mathematics, University of California, Santa Barbara, California 93106-6105, USA}

\author{Lu Wei}
\affiliation{School of Gifted Young and School of Physics, University of Science and Technology of China, Hefei, Anhui, 230026, P.R. China}

\author{Yu-Chun Wu}
\email{\tt wuyuchun@ustc.edu.cn}
\affiliation{CAS Key Laboratory of Quantum Information, School of Physics, University of Science and Technology of China, Hefei, Anhui, 230026, P.R. China}
\affiliation{CAS Center For Excellence in Quantum Information and Quantum Physics, University of Science and Technology of China, Hefei, Anhui, 230026, P.R. China}

\author{Guang-Can Guo}
\affiliation{CAS Key Laboratory of Quantum Information, School of Physics, University of Science and Technology of China, Hefei, Anhui, 230026, P.R. China}
\affiliation{CAS Center For Excellence in Quantum Information and Quantum Physics, University of Science and Technology of China, Hefei, Anhui, 230026, P.R. China}

\author{Guo-Ping Guo}
\affiliation{CAS Key Laboratory of Quantum Information, School of Physics, University of Science and Technology of China, Hefei, Anhui, 230026, P.R. China}
\affiliation{CAS Center For Excellence in Quantum Information and Quantum Physics, University of Science and Technology of China, Hefei, Anhui, 230026, P.R. China}
\affiliation{Origin Quantum Computing Hefei, Anhui 230026, P. R. China}

\begin{abstract}
A study of the artificial neural network representation of quantum many-body states is presented.
The locality and entanglement properties of states for shallow and deep quantum neural networks are investigated in detail.
By introducing the notion of local quasi-product states,  for which the locally connected shallow feed-forward neural network states and restricted Boltzmann machine states are special cases,
we show that R\'{e}nyi entanglement entropies of all these states obey the entanglement area law.
Besides, we also investigate the entanglement features of deep Boltzmann machine states and show that locality constraints imposed on the neural networks make the states obey the entanglement area law.
Finally, as an application, we apply the notion of R\'{e}nyi entanglement entropy to understand the power of neural networks, and show that image classification problems can be efficiently solved must obey the area law.
\end{abstract}
\maketitle

\section{Introduction}
\label{sec:RBM}
Understanding entanglement features of quantum systems is crucial for understanding many important physical phenomena.
One of the most outstanding issues of entanglement features is that entanglement entropy is somehow bounded by the area of quantum systems.
This idea, now as an important part of holography principle, can be applied into many different physical areas,
such as topological order \cite{wen1995topological,KITAEV2006,KONG201762}, fractional quantum hall effect \cite{Cappelli_2018},
topological insulator and topological superconductor \cite{Hasan2010,Qi2011RMP}, anti-de Sitter Space/Conformal field theory (AdS/CFT) correspondence \cite{THOOFT1985727,Susskind1995,witten1998anti,Bousso2002} and so on.

Holography principle
asserts there is a duality between the boundary quantum field theory and the bulk gravitational theory.
More precisely, it claims that the $d+1$ dimensional conformal field theories ($\mathrm{CFT}_{d+1}$) are equivalent to the gravitational theory on $d+2$ dimensional anti-de Sitter space $\mathrm{AdS}_{d+2}$.
Based on the holographic approach, Ryu and Takayanagi proved that the entanglement entropy of a subsystem $\mathcal{A}$ in $\mathrm{CFT}_{d+1}$ is related to the area of static minimal surface $\gamma_{\mathcal{A}}$ in $\mathrm{AdS}_{d+2}$ whose boundary matches the boundary $\partial \mathcal{A}$,
the famous Ryu-Takayanagi formula \cite{Ryu2006} reads
$$S(\mathcal{A})=\frac{\mathrm{Area}(\gamma_{\mathcal{A}})}{4G_{N}^{(d+2)}},$$
where $G_{N}^{(d+2)}$ is the $d+2$-dimensional Newton constant. The key point here is that the entanglement entropy is bounded by the area of the quantum system and there is a duality between geometry and entanglement.
For applications in quantum many-body systems, it is now a well-known result that the ground states of local gapped quantum systems obey the entanglement area law \cite{Eisert2010,hartnoll2018holographic}:
the value of entanglement entropy between a subsystem $\mathcal{A}$ and its complement $\mathcal{A}^c$ scales at most the area $\mathrm{Area}(\mathcal{A})$ rather than the volume $\mathrm{Vol}(\mathcal{A})$ of subsystem $\mathcal{A}$.
It can be understood intuitively that the entanglement area law is a result of the fact that the correlations of particles in a ``natural'' quantum system are usually local,
thus the contribution to the entanglement entropy between $\mathcal{A}$ and $\mathcal{A}^c$ given by cutting the correlated pairs between  $\mathcal{A}$ and $\mathcal{A}^c$ only depends on the pairs of particles in the vicinity of the boundary.
Although there are many numerical and theoretical results support this intuitive argument mainly in $(1+1)$ system and in some $(2+1)D$ systems,
rigorously proving the entanglement area law is extremely challenging and many sophisticated mathematical tools,
like Toeplitz matrix theory \cite{bottcher2000toeplitz,bottcher2013analysis}, Fisher-Hartwig theorem \cite{basor1978asymptotic}, Lieb-Robinson bound \cite{lieb1972finite}, Chebyshev polynomial \cite{arad2013area} and so on, must be used.
It is now one of the central problems in Hamiltonian complexity theory  to establish the entanglement area law.

On the other hand, given a real quantum many-body system, we will be facing an extremely large (about $10^{23}$ or more) degrees of freedom,
which makes it a notoriously difficult task to solve the Schr\"{o}dinger equations directly. However, fortunately, physical systems often have a  simplified internal structure,
for which we can use exponentially fewer parameters to characterize the ground states and time evolutions of the system,  this makes many numerical and theoretical methods possible.
The traditional mean-field approach can solve the equations for many weakly correlated systems. For strongly correlated quantum systems, many new tools are developed these years.
Quantum Monte Carlo sampling \cite{Foulkes2001quantum} provides a high-accuracy method for studying large systems, however, it suffers from the sign problem which makes it unable to be applied to frustrated spin systems and interacting fermion systems.
Tensor network representation of quantum states, such as density-matrix renormalization group (DMRG) and matrix product states \cite{White1992}, projected entangled pair states (PEPS) \cite{verstraete2004renormalization}, folding algorithm \cite{Banuls2009},
entanglement renormalization \cite{Vidal2007}, time-evolving block decimation (TEBD) \cite{Vidal2003}, etc., play an important role in calculating $1d$ and $2d$ quantum systems and even in the construction of AdS/CFT correspondence \cite{Pastawski2015,almheiri2015bulk}.
Among all of these numerical and theoretical methods to represent and approximate quantum states,
the neural network as an important tool of machine learning, which shows great power in approximating given functions and extracting features from a big set of data, is now attracting many interests from both physicists and computer scientists.

Neural networks are recently introduced as a new representation of quantum many-body states \cite{Carleo602}, and it shows great potential in solving some traditionally difficult quantum problems,
for instance,  solving some physical models and studying the time evolution of these systems \cite{Carleo602,Deng2017}, representing toric code states \cite{Deng2017a}, graph states \cite{gao2017efficient},
stabilizer code  states\cite{jia2018efficient,zhang2018efficient} and topologically ordered states \cite{huang2017neural,Deng2017a,jia2018efficient,Lu2019efficient}, studying quantum tomography \cite{Zhang2017,torlai2017many}, and so on.
Quantum neural network states are currently subject to intense research and represents a new direction for efficiently calculating ground states and unitary evolutions of many-body quantum systems.
These researches stimulate an explosion of results to apply machine learning methods to investigate condensed matter physics,
like distinguishing phases \cite{carrasquilla2017machine}, quantum control \cite{August2017}, error-correcting of topological codes \cite{Torlai2017}, etc. The interplay between machine learning and quantum physics has given birth to  a new discipline, now known as quantum machine learning.

In this work, we present a study of the entanglement properties of the quantum neural network state. It has been shown that locally connected restricted Boltzmann states obey the entanglement area law \cite{Deng2017}.
Here we give a more comprehensive study of the entanglement properties of both shallow and deep neural network states.
And as an application, we apply the notion of entanglement entropy to the understanding of the representational powers of neural networks in image classification problems.

The paper is organized as follows. In Sec. \ref{sec:shallow}, we introduce the notion of local quasi-product state and establish the entanglement area law of these states. Since locally connected neural network states are special cases of the local quasi-product states, they also obey the entanglement area law. Sec. \ref{sec:DBM} presents the study of the deep Boltzmann machine (DBM) states, by introducing the geometry of deep Boltzmann machine, we prove that local DBM states obey the entanglement area law. In Sec. \ref{sec:image}, we apply the notion of R\'{e}nyi entropy to the understanding of the power of the neural network in solving image classification problem, and we show that the target function of classification problem of local smooth images obeys the entanglement area law.  Finally, we discuss in the last section some subtle issues of the locality and entanglement of the neural network states.

\section{Area-law entanglement of local quasi-product states and its applications to shallow neural network states}
\label{sec:shallow}

\subsection{Notion of quasi-product states}
The Schr\"{o}dinger equation of condensed matter system usually involves a large number of degrees of freedom which makes it extremely difficult to be solved exactly.  However, the eigenstates of the Hamiltonians of these \emph{natural} systems often have an internal simplified structure, which makes many approximating or even exact methods possible. Neural network states are introduced as ansatz states of many-body quantum systems recently, and because of their good performance in solving some problems which can not be solved using the state-of-the-art method, many attentions are attracted \cite{Carleo602,Deng2017,gao2017efficient,Jia2019quantum}. Here, to explore the area-law entanglement of the neural network states, we first introduce the concept of quasi-product states. As we will show later, the locality constraint imposed on the neural network architecture results in states of quasi-product form.

Let $\mathcal{S}=\{s_1,\cdots,s_N\}$ be a system with $N$ particles, by a local $K$-cluster cover we mean a class of local subsets of $\mathcal{S}$, viz., $\mathcal{C}_1,\cdots, \mathcal{C}_M$, called local cluster, for which each $\mathcal{C}_i$ only contain at most $K$ particles in a local region and $\cup_{i=1}^M \mathcal{C}_i=\mathcal{S}$. A local $K$-cluster quasi-product state can then be defined as $\Psi(s_1,\cdots,s_N)=\Phi_1(\mathcal{C}_1)\times \cdots \times \Phi_M(\mathcal{C}_M)$, where each cluster term $\Phi_i(\mathcal{C}_i)$ is a function of degrees of freedom of particles contained in $\mathcal{C}_i$, and the size of clusters $K:=\max\{|\mathcal{C}_i|\}$ does not depend on the system size $N$. It is obvious that product state is just $1$-local quasi-product state, i.e., each $\Phi_i(\mathcal{C}_i)$ is just the function $\Phi_i(s_i)$, since each local cluster $\mathcal{C}_i$ only contains one particle $s_i$, we will also refer this kind of states as local 1-cluster quasi-product state.


It turns out that many crucial classes of quantum states can be expressed as local quasi-product states, such as cluster state, $\mathbb{Z}_2$-toric code states,  graph states, $\mathbb{Z}_2$-stabilizer code states, Kitaev's $D(\mathbb{Z}_d)$ quantum double ground states.  They are all explicitly constructed in local RBM form \cite{Deng2017a,gao2017efficient,jia2018efficient,zhang2018efficient}, but we will show later in this section that all local RBM states are the local quasi-product state.

Many of examples of local  gapped systems come from local commutative Hamiltonian $H=\sum_{k}H_k$ for which $[H_k,H_l]=0,\forall k,l$ and each local term $H_k$ only acts on a local region $\mathcal{S}_k$ of the system\footnote{In general, we can regard the lattice of the system as a graph $\mathcal{S}$, for each vertex $v$, there is a corresponding local Hilbert space $\mathcal{H}_v$. The total space is then $\mathcal{H}_{tot}=\otimes_{v\in \mathcal{S}}$. In this given graph, we can define the $r$-range neighborhood of a particle $v$ as the set of all particles which are at most $r$-path far from $v$. In this way, we can define the background geometry of the system.}. It is very natural to use the quasi-product state as an ansatz state to solve the eigenvalue equation $H\Psi(s_1,\cdots,s_N)=E_0\Psi(s_1,\cdots,s_N)$. We can assign a cluster $\mathcal{C}_k$ to each local term $H_k$ and usually  we also make constraint that $\mathcal{S}_k\subseteq \mathcal{C}_k$, i.e., $\mathcal{C}_k$ contain all particles which $H_k$ acts on nontrivially. In this way, the eigenvalue equation can be simplified as
$$H_k\prod_{j:\mathcal{C}_j\cap \mathcal{S}_k\neq \emptyset}\Phi_j(\mathcal{C}_j)=E_{0}^{(k)}\prod_{j:\mathcal{C}_j\cap \mathcal{S}_k\neq \emptyset}\Phi_j(\mathcal{C}_j),$$
where $E_{0}^{(k)}$ is the ground state energy of $H_k$. Then using some other properties, like symmetry, of the system, we can alternatively solve these equations with less variables to give the solution of the original eigenvalue problem. Here, for illustration, we choose cluster stabilizer code, toric code and graph state as examples.

\begin{example}
Cluster stabilizer code state, or equivalently, the ground state of $(1+1)D$ symmetry protected topological (SPT) phase Hamiltonian
$$H_{\mathrm{cluster}}=\sum_{k=1}^N (I- \sigma_{k-1}^z\sigma_{k}^x\sigma_{k+1}^z)$$
defined on a $1d$ lattice with periodic boundary condition can be represented by a $3$-local quasi-product state. Each term $\sigma_{k-1}^z\sigma_{k}^x\sigma_{k+1}^z$ is called a stabilizer.  The cluster state is a $\mathbb{Z}_2\times\mathbb{Z}_2$ protected topological state \cite{Son2012}, which can used for measurement-based quantum computation \cite{Briegel2001persistent,Raussendorf2003,NIELSEN2006}. Here, we validate the efficiency of the local quasi-product state representation by explicit constructions, the local cluster is chosen as 3-local cluster corresponding to each stabilizer, i.e., $\Phi_k(\mathcal{C}_k)=\Phi_k(s_{k-1},s_k,s_{k+1})$. The ground state satisfies
$\sigma_{k-1}^z\sigma_{k}^x\sigma_{k+1}^z\sum_{s_1,\cdots,s_N=\pm 1}\Psi(s_1,\cdots,s_N)|s_1,\cdots,s_N\rangle=\sum_{s_1,\cdots,s_N=\pm 1}\Psi(s_1,\cdots,s_N)|s_1,\cdots,s_N\rangle$ for all $k$, which is equivalent to
\begin{align}
  &s_{k-1}s_{k+1}\Psi(\cdots,s_{k-1}, s_{k},s_{k+1},\cdots)\nonumber\\
  =&\Psi(\cdots,s_{k-1}, -s_{k},s_{k+1},\cdots), \forall k .
\label{eq:clusterstate}
\end{align}
Using the 3-local quasi-product state ansatz state $\Psi(s_1,\cdots,s_N)=\prod_{k=1}^N \Phi_{k}(s_{k-1},s_k,s_{k+1})$ and cancelling the same term from two sides of the equality, we obtain
\begin{align}
& s_{k-1}s_{k+1} \Phi_{k-1}(s_{k-2},s_{k-1},s_k)\times \Phi_{k}(s_{k-1},s_{k},s_{k+1}) \nonumber\\
 &\times \Phi_{k+1}(s_{k},s_{k+1},s_{k+2})  = \Phi_{k-1}(s_{k-2},s_{k-1},-s_k)\nonumber \\
& \times \Phi_{k}(s_{k-1},-s_{k},s_{k+1}) \times \Phi_{k+1}(-s_{k},s_{k+1},s_{k+2}), \forall k.\nonumber
\label{}
\end{align}
These are highly nonlinear equations, thus are very difficult to solve directly and the solution is not unique in general. But noticing that the model is translationally invariant, we can assume that all local clusters $\Phi_k(\mathcal{C}_k)$ are of the same form. Via this simplification, we can obtain a solution:
$$\Phi_k(s_{k-1},s_k,s_{k+1})=2\cos (\frac{\pi+2\pi s_{k-1}+3\pi s_k+\pi s_{k+1}}{4}).$$
It is easily verified that the local quasi-product state $\Psi(s_1,\cdots,s_N)=\prod_{k=1}^N \Phi_{k}(s_{k-1},s_k, s_{k+1})$ satisfies the Eq. (\ref{eq:clusterstate}) .
\end{example}

\begin{example}
\label{example:toric}
\begin{figure}
  \centering
  \includegraphics[width=8cm]{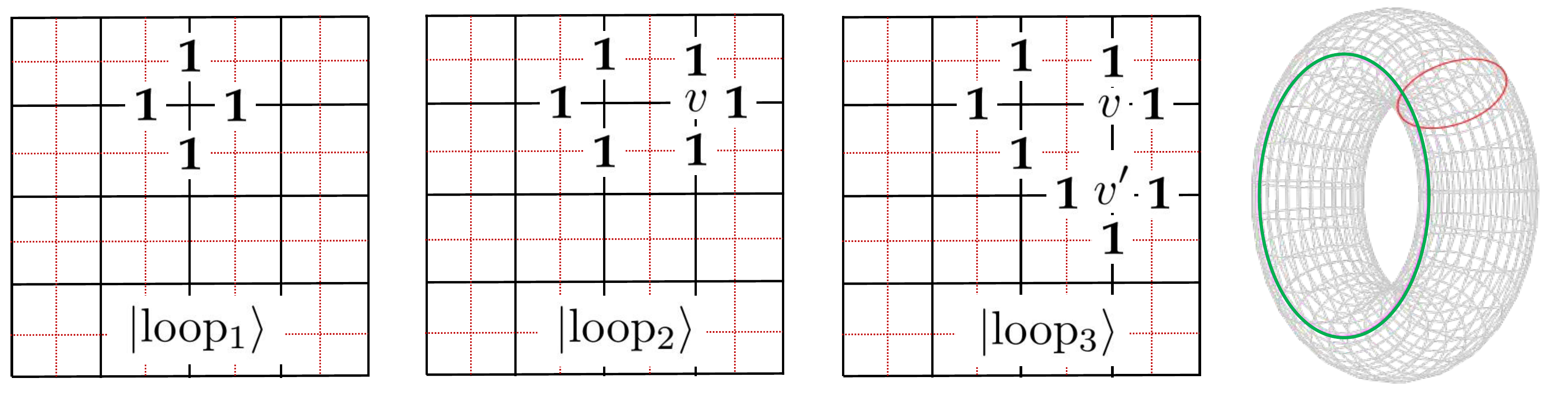}\\
  \caption{The loop spin configurations in dual lattice which is represented as the red dash lines, the unlabeled edges are all set to be in $|0\rangle$. The vertex operator can transform one loop into the other, e.g., $A_v|\mathrm{loop_1}\rangle=|\mathrm{loop_2}\rangle$ and $A_{v'}A_v|\mathrm{loop_1}\rangle=A_{v'}|\mathrm{loop_2}\rangle=|\mathrm{loop_3}\rangle$. For the torus case, longitudinal loop (red one) and latitudinal loop (green one) are essentially different kinds of loops, they can not be deformed into each other by vertex operators.}\label{fig:GSDloop}
\end{figure}

Let us now consider the toric code model ($\mathbb{Z}_2$-Kitaev quantum double model) \cite{Kitaev2003}, which is the simplest model of topologically ordered states and plays an important role in quantum error correcting codes and fault-tolerant quantum computation theories \cite{Nayak2008}. Given an $L\times L$ square lattice with periodic boundary condition (i.e., on a $2d$ torus $\mathbb{T}^2$), on each edge there is an associated spin space $\mathbb{C}^2$, thus there will be $N= 2L\times L$ qubits in total. To each vertex $v$ and plaquette we assign a stabilizer operator $A_v=\prod_{j\in \partial v}\sigma^x_j$ and $B_p=\prod_{j\in \partial p}\sigma^z_j$ respectively, and the Hamiltonian is of the form
$$H_{\mathrm{toric}}=-\sum_vA_v-\sum_pB_p.$$
The ground state of the Hamiltonian is four-fold degenerate (which corresponds to the order of the first $\mathbb{Z}_2$ homology group of torus $\mathbb{T}^2$, viz., $GSD=|H_1(\mathbb{T}^2,\mathbb{Z}_2)|$ ). Let us briefly recall how to calculate the ground state of the toric code model. Consider the constraints imposed by plaquette operators, i.e., $B_p|\Omega\rangle=|\Omega\rangle$. In the $\sigma^z$ basis $\{|0\rangle, |1\rangle\}$, the set of spin configurations is then $\{|00\cdots 0\rangle, |00\cdots 1\rangle,\cdots,|11\cdots 1\rangle\}$. Assume that
\begin{equation}\label{eq:GSD}
|\Omega\rangle=\sum_{\mathbf{s}}c_{\mathbf{s}}|\mathbf{s}\rangle,\nonumber
\end{equation}
then for $B_p=\sigma^z_{p_1}\sigma^z_{p_2}\sigma^z_{p_3}\sigma^z_{p_4}$, we have $B_p|\mathbf{s}\rangle =(-1)^{s_{p_1}+s_{p_2}+s_{p_3}+s_{p_4}}|\mathbf{s}\rangle$,
where $|\mathbf{s}\rangle=|s_{p_1}s_{p_2}s_{p_3}s_{p_4}\rangle\otimes|\cdots\rangle$ and the addition is modulo two. Therefore 
$$B_p|\Omega\rangle =\sum_{\mathbf{s}}c_{\mathbf{s}}(-1)^{s_{p_1}+s_{p_2}+s_{p_3}+s_{p_4}}|\mathbf{s}\rangle=\sum_{\mathbf{s}}c_{\mathbf{s}}|\mathbf{s}\rangle.$$ 
If $s_{p_1}+s_{p_2}+s_{p_3}+s_{p_4}=1$ the corresponding coefficient $c_{\mathbf{s}}$ must be zero, thus the spin configurations which do not vanish is the one only even number of $|1\rangle$ is placed on the edges of each plaquette, which means that the $|1\rangle$ spins form a close loop in the dual lattice (see Fig. \ref{fig:GSDloop}):
\begin{equation}\label{eq:GSDD}
|\Omega\rangle=\sum_{\mathrm{loop}}c_{\mathrm{loop}}|\mathrm{loop}\rangle.\nonumber
\end{equation}

As shown in Fig. \ref{fig:GSDloop}, the effect of vertex operators is just to deform one loop into another. We call two loops equivalent if they can be linked by some vertex operators, the GSD is just the number of equivalent classes of loops. There exist essentially four different kinds of loops like longitudinal loop and latitudinal loop shown in Fig. \ref{fig:GSDloop}, they are actually the bases (logical states) of the ground state space of toric code model:
\begin{gather}
|00\rangle_L=\sum_{\text{trivial loop}}|\text{loop}\rangle,\nonumber\\
|01\rangle_L=\sum_{\text{longitudinal loop}}|\text{loop}\rangle,\nonumber\\
|10\rangle_L=\sum_{\text{latitudinal loop}}|\text{loop}\rangle,\nonumber\\
|11\rangle_L=\sum_{\text{longitudinal+latitudinal loop}}|\text{loop}\rangle.\nonumber
\end{gather}

Here, by explicit construction, we show that the ground state of toric code model can be represented as $4$-local quasi-product state, the philosophy is very similar as what we have done above. Actually, we can assign a cluster to each vertex and plaquette, thus the state  is  of the form $\Psi(s_1,\cdots,s_N)=\prod_v \Phi_v(\mathcal{C}_v)\prod_p\Phi_p(\mathcal{C}_p)$, where $\mathcal{C}_v$ (resp. $\mathcal{C}_p$) only contain spins which $A_v$ (resp. $B_p$) acts nontrivially.
Like in the cluster state construction, we have the constraints
\begin{align}
&A_v\Psi(s_1,\cdots,s_N)=\Psi(\cdots, -s_{j_1}^{(v)},-s_{j_2}^{(v)},-s_{j_3}^{(v)},-s_{j_4}^{(v)},\cdots)\nonumber\\
&B_p \Psi(s_1,\cdots,s_N)=\prod_{i\in\partial p}s_i\Psi(s_1,\cdots,s_N).\nonumber
\end{align}
This can be transformed into a set of equations only involves local clusters around vertex $v$ and plaquette $p$, as we have show exlicitly in  Refs. \cite{jia2018efficient,zhang2018efficient} for general stabilizer code. A solution in the RBM form in provided in \cite{Deng2017a}, it is easily checked that the corresponding local clusters 
\begin{align}
& \Phi_v(s_{i_1}^{(v)},s_{i_2}^{(v)},s_{i_3}^{(v)},s_{i_4}^{(v)})=\cos[\frac{\pi}{2}(s_{i_1}^{(v)}+s_{i_2}^{(v)}+s_{i_3}^{(v)}+s_{i_4}^{(v)})],\nonumber\\
& \Phi_p(s_{j_1}^{(p)},s_{j_2}^{(p)},s_{j_3}^{(p)},s_{j_4}^{(p)}) =\cos[\frac{\pi}{4}(s_{j_1}^{(p)}+s_{j_2}^{(p)}+s_{j_3}^{(p)}+s_{j_4}^{(p)})],\nonumber
\end{align}
form a ground state of the toric code model. The excited state can also be represented in quasi-product form in a similar way. We must stress here that this is just one of the solutions, in fact there are a lot of other solutions depending on the choice of the local clusters.
\end{example}

\begin{example}
Another example we will consider here is the graph state , which is an important class of multipartite entangled quantum states and is useful for quantum error correcting codes, measurement-only quantum computation and so on \cite{hein2006entanglement}. For a given graph $G$ with vertex set $V(G)=\{1,\cdots,N\}$ and edge set $E(G)\subset V(G)\times V(G)$, the graph state is defined as
$$|\Psi_G\rangle=\prod_{\langle ij\rangle \in E(G)}U_{\langle ij\rangle}|+\rangle^{\otimes V(G)},$$
where $U_{\langle ij\rangle}$ is a two-qubit controlled-$Z$ gate. The wave function thus takes the form
\begin{equation}
\Psi_G(s_1,\cdots,s_N)=\prod_{\langle ij\rangle\in E(G)}\frac{(-1)^{s_is_j}}{\sqrt{2}}.
\label{eq:graph}
\end{equation}
To represent the state using the local quasi-product state, we can assign a cluster to each edge $e=\langle i_e j_e\rangle \in E(G)$, i.e.,
$\Psi_G(s_1,\cdots,s_N)=\prod_{e\in E(G)}\Phi_e(s_{i_e},s_{j_e})$.
From Eq. (\ref{eq:graph}) we see that $\Phi_e(s_{i_e},s_{j_e})=\frac{(-1)^{s_{i_e}s_{j_e}}}{\sqrt{2}}$,  which is obviously a $2$-local quasi-product state.
\end{example}

\subsection{Shallow neural network states}
Here we will construct two important classes of quasi-product states via feed-forward and stochastic recurrent neural networks, which will be the main focus of this work. To this end, we first need to introduce the notion of geometry for neural networks.

\subsubsection{The geometry of neural network states}
\begin{figure}
  \centering
    \includegraphics[width=8.5cm]{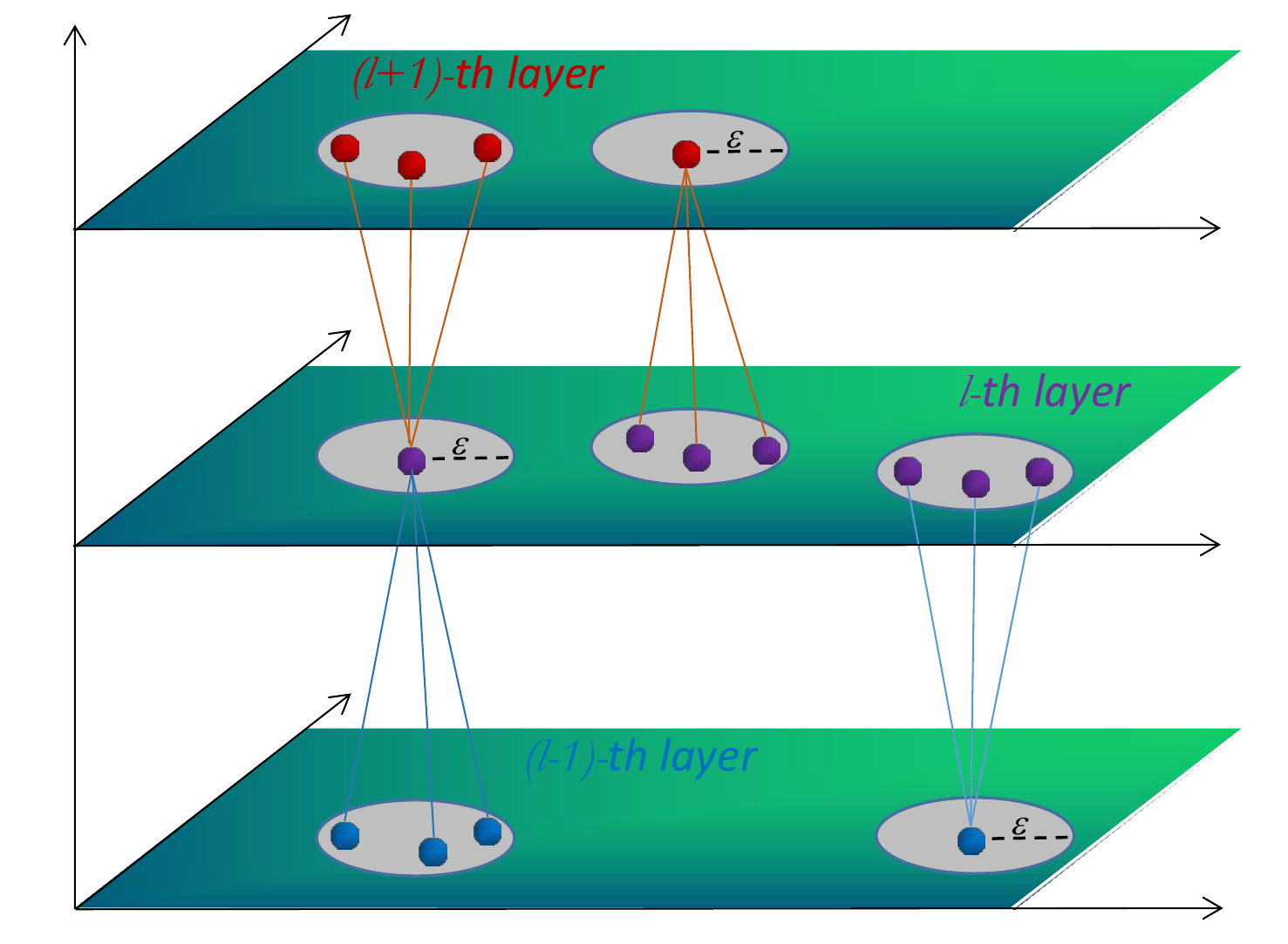}\\
  \caption{The depiction of the $K$-local neural networks, where the geometries of each layer are the same and all are duplicated from the geometry of the physical layer. For each neuron $h$ we can define the $\varepsilon$-neighborhood $B(h; \varepsilon)$, $h$ can only connect neurons lie in $B(h; \varepsilon)$ in the previous and the next. Under this local constraint, the maximum number $K$ in the given  $\varepsilon$-connected neural network is defined as the maximum of the number of neurons that can be connected. Here is an example of $3$-local neural network architecture.}\label{fig:layer}
\end{figure}
Inspired by the geometry of tensor network states \cite{ORUS2014}, here we introduce the notion of the geometry of the neural network states, which turns out to be crucial for understanding entanglement features. Hereinafter, we will concentrate on the neural networks with a layered structure, which are also the most studied cases. The physical degrees of freedom are placed on some fixed layer of neural networks, e.g., the visible layer of restricted Boltzmann machine or input layer of the feed-forward neural network, the layer will be referred to as \emph{physical layer}. Notice that the physical layer has its geometry given by the physical system. For example, if the physical degrees of freedom (like spins) are placed on the square lattice, we can impose the neurons (represent physical degrees of freedom) to have the same square lattice geometry.  After the geometry of the physical layer is fixed by the geometry of the physical system, all other layers are imposed to have the same geometry duplicated from the physical layer as shown in Fig. \ref{fig:layer}.

Recall that the geometry of tensor network states is characterized by the positions of local tensors and their contraction pattern, here for neural network states, similar results hold. We can compare the distance between neurons in different layers since these layers have the same geometry. Now we can define the notion of locality of a neural network. For a  neuron $h_i^{(l)}$ in a given, say $l$-th,  layer, it is called local $\varepsilon$-connected with other neurons if it only connects to the neurons in $(l-1)$-th and $(
l+1)$-th layers in the $\varepsilon$-neighborhood of $h_i$ (see Fig. \ref{fig:layer} for illustration). If all the neurons of a neural network are local $\varepsilon$-connected with each other, we say the neural network is a local $\varepsilon$-connected neural network. Similar construction has been used in Refs \cite{Deng2017a,jia2018efficient,zhang2018efficient,You2018} for exactly constructing neural network states of some physical systems. When the in a  local $\varepsilon$-connected neural network, each neuron $h^{(l)}_i$ only connects with $K$ neurons both in  $(l-1)$-th and $(
l+1)$-th layers, we call it a $K$-local neural network.  For a $K$-local neural network, a corresponding quantum state can be given. Usually, there are two different ways to build quantum neural network states \cite{Jia2019quantum}, the first approach, which is also the approach we choose to use in this work, is to introduce complex weights and biases into the neural network; the second approach is to represent the amplitude and phase of a wavefunction separately.  We will prove that quantum states build from $K$-local neural networks obey the entanglement area-law, since there are all quasi-product states, and the entanglement area law of quasi-product states will be established later. To make this construction more clear, let us see two important examples.

\subsubsection{Local restricted Boltzmann machine states}

In this part, we will introduce the notion of restricted Boltzmann machine states, which were introduced in Ref. \cite{Carleo602} for calculating ground state and unitary evolution of strongly correlated many-body systems. The RBM was invented by Smolensky~\cite{smolensky1986information}, it is an energy-based neural network model \cite{hinton1983optimal,ackley1985learning}. Since RBM only has two layers of neurons, one visible layer and one hidden layer, it can be regarded as a shallow neural network.

We now build quantum states from local RBM and show that they are local quasi-product states. The RBMs have a layered structure, which makes the locality defined above applicable. To construct a local RBM state, we first impose the locality constraints on the visible layer, which is nothing but the physical layer, each visible neuron corresponds to the physical degrees of freedom (e.g. spin), denoted as $\mathcal{S}=\{v_1,\cdots,v_n\}$, and the geometry of the visible layer is inherited from the physical system. The hidden neurons are denoted as $\{h_1,\cdots,h_m\}$, which are placed on the hidden layer with geometry duplicated from visible layer, viz., the distance between two neurons can be defined the same as visible layer (the distance of visible layer is inherited from the physical system). The weight between $h_j$ and $v_i$ is denoted as $W_{ij}$, the biases of $v_i$ and $h_j$ are $a_i$ and $b_j$ respectively. The RBM representation of quantum states is obtained by tracing out all hidden neurons, viz.,
\begin{align}\label{eq:localrbm}
  &\Psi_{RBM}(v_1,\cdots,v_n)\nonumber\\
  =&\sum_{h_1,\cdots,h_m} \exp\{\sum_ia_iv_i+\sum_jh_j(b_j+\sum_{i:\langle ij\rangle}W_{ij}v_i)\}\nonumber\\
  =&e^{\sum_ia_iv_i}\prod_j\Gamma_{j}(v_i:\langle v_i h_j\rangle),
\end{align}
where $\Gamma_j(v_i:\langle v_ih_j\rangle)=2\cosh(b_j+\sum_{i:\langle ij\rangle}v_iW_{ij})$ if $h_j=\pm 1$ and $\Gamma_j(v_i:\langle v_i h_j\rangle)=1+\exp(b_j+\sum_{i:\langle ij\rangle}v_iW_{ij})$ if $h_j=0,1$, and by notation $\langle ij\rangle$ we mean $h_j$ and $v_i$ are connected. The $K$-local RBM state can be defined as the one whose hidden neurons $h_j$ only connect with at most $K$ visible neurons $v_{j_k}$ in a local $\varepsilon$-neighborhood of $h_j$. From the construction it is easily checked that the $\Psi_{RBM}(v_1,\cdots,v_n)=\prod_j\Phi_j(v_i: \langle v_i h_j\rangle)$. This kind of construction has be used in Refs \cite{Deng2017a,jia2018efficient,zhang2018efficient,You2018} for investigating physical properties of complex systems.

\begin{figure}
  \centering
 \includegraphics[width=8cm]{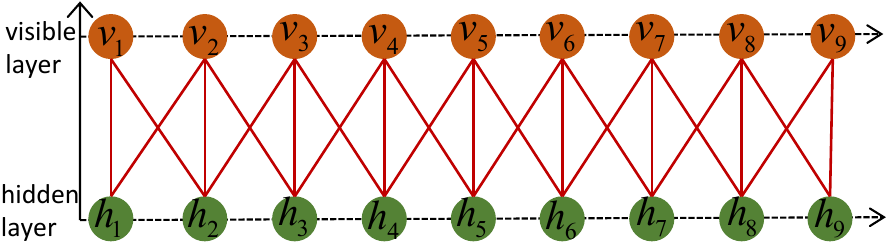}\\ (a) local RBM states.\\
 \includegraphics[width=8cm]{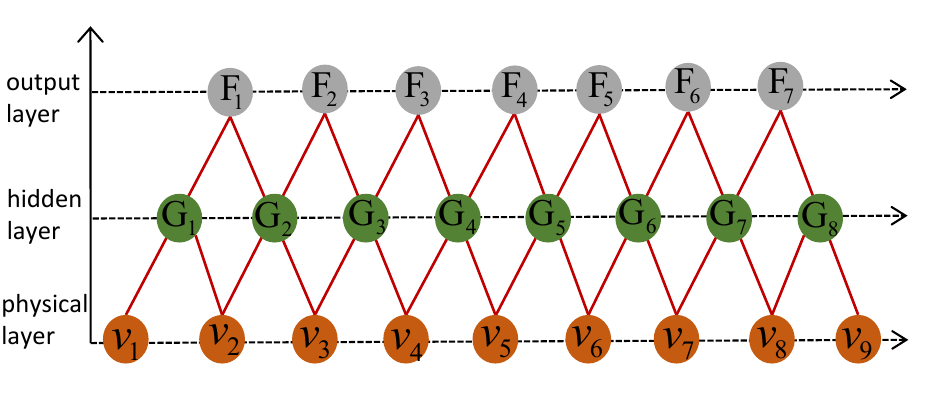}\\  (b) local feed-forward neural network states.
  \caption{(a) Example of one-dimensional $3$-local RBM state; (b) example of one-dimensional $3$-local feed-forward neural network state.}\label{fig:localexample}
\end{figure}

Let us now see an one-dimensional example. As shown in Fig. \ref{fig:localexample} (a), the quantum states built from $3$-local RBM neural network is a local quasi-product state. From Eq. (\ref{eq:localrbm}), it is easily checked that $\Psi(v_1,\cdots,v_9)=\Phi_1(v_1,v_2)\times\Phi_{2}(v_1,v_2,v_3)\times\cdots\times\Phi_{9}(v_8,v_9)$, thus it is a $3$-local quasi-product state.

\subsubsection{Local feed-forward neural network states}
Another crucial class of quasi-product states is local feed-forward neural network states.
To start with, let us first briefly recall the notion of a feed-forward neural network. The neuron of the feed-forward neural network is modeled by McCulloch-Pitts neuron model \cite{McCulloch1943}, $n$ inputs $x_1,x_2,\cdots,x_n$ values are transmitted by $n$ corresponding weighted connections with weights $w_1,w_2,\cdots,w_n$. After the input values have reached the neuron, they are added together with weights $\sum_{i=1}^n w_ix_i$ and the result is then compared with the bias $b$ of the neuron to determine if it is activated or deactivated. The activation status is characterized by the activation function $F$. Therefore the output of the neuron is given by $y=F(\sum_{i=1}^n w_ix_i-b)$. There are several commonly used activation functions such as step function, sigmoid function and so on. Here, to make the construction more general, we won't restrict the form of activation function and we allow the activation function of each neuron to be different in one neural network. A feed-forward neural network is several layers of neurons for which the neurons in adjacent layers are connected with each other but there is no intra-layer connection, as shown in Fig. \ref{fig:localexample} (b).

To build quantum states from feed-forward neural network, complex weights, biases and complex activation functions need to be introduced \footnote{In practical applications, introducing complex weights and biases and complex activation functions may lead to some difficulties for training the neural network, to overcome this shortage, the amplitude and phase of quantum state are usually represented by two feed-forward neural network separately, see Ref, e.g., \cite{Jia2019quantum}. But here for our purpose, we choose to use the complex neural network approach.}. We assume the output value of the output layer if $y_1=F_1(v_1,\cdots,v_n),\cdots,y_m=F(v_1,\cdots,v_n)$, the quantum states is construct as their product, $\Psi(v_1,\cdots,v_n)=\prod_{j=1}^{m}F_j(v_1,\cdots,v_n)$ where the normalization factor is omitted. If we add the locality constraints of connections between each layer, then we get the local feed-forward neural network states. See Fig. \ref{fig:localexample} (b) for an example. The local constraints make the corresponding states quasi-product states. As in Fig. \ref{fig:localexample} (b), the value of the first output neuron $\Phi_1(v_1,v_2,v_3)=F[G_1(v_1,v_2),G_2(v_2,v_3)]$ only depends on particles $v_1,v_2,v_3$, the corresponding quantum states is of the form $\Psi(v_1,\cdots,v_9)=\Phi_1(v_1,v_2,v_3)\times \cdots \times \Phi_7(v_7,v_8,v_9)$ which is obviously a quasi-product state. It worth mentioning that the number of layers of network should not be too large, otherwise the size of local cluster $|\mathcal{C}|$ of the corresponding states will be comparable with the system size $N$, which will break the locality constraint.

\subsection{Entanglement area law of local quasi-product states}

\begin{figure}
  \centering
    \includegraphics[width=7cm]{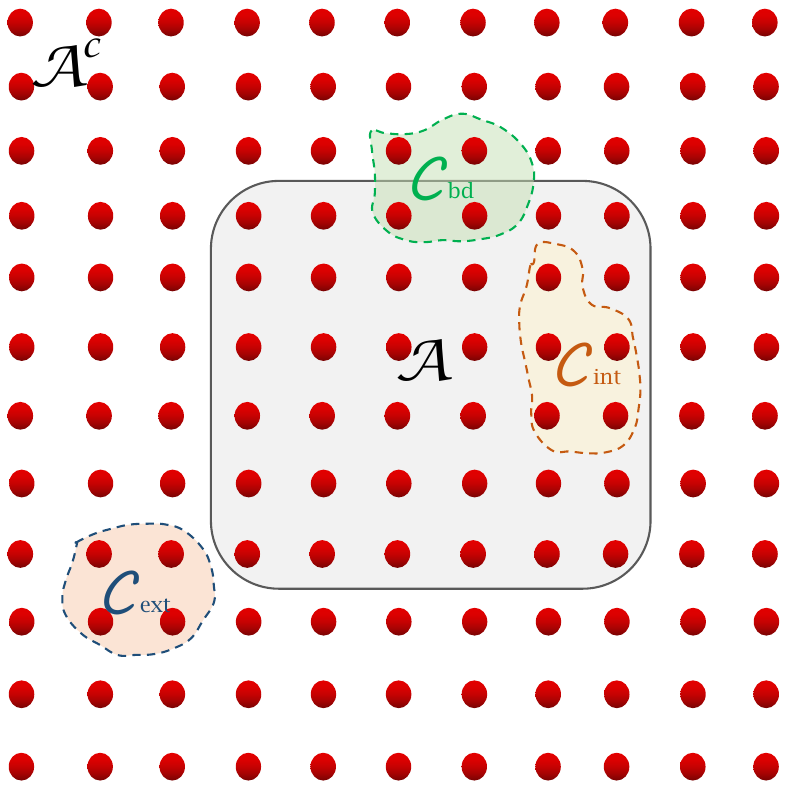}\\
  \caption{The depiction of internal cluster $\mathcal{C}_{int}$, external cluster $\mathcal{C}_{ext}$ and boundary cluster $\mathcal{C}_{bd}$ for a bipartition $\mathcal{A}$ and $\mathcal{A}^c$ of the system.}\label{fig:entropy}
\end{figure}

Entanglement entropy is a crucial theoretical tool for investigating quantum many-body systems. We now establish the entanglement area law of local quasi-product states, since local RBM states and local feed-forward neural network states are special cases of local quasi-product states, they all obey the entanglement area law. We prove that in arbitrary spatial dimensions the local quasi-product states obey the entanglement area law for arbitrary connected bipartition of the system. More precisely, we have the following theorem:
\begin{thm}For an N-particle system $\mathcal{S}$, suppose that $$|\Psi\rangle=\sum_{s_1,\cdots,s_N}\Psi(s_1,\cdots,s_N)|s_1\rangle\otimes\cdots\otimes|s_N\rangle$$ is a $K$-local quasi-product quantum states, then the R\'{e}nyi entropies of the reduced density matrix $\rho_{\mathcal{A}}=\mathrm{Tr}_{\mathcal{A}^c}|\Psi\rangle\langle\Psi|$ with respect to the bipartition $\mathcal{S}=\mathcal{A}\sqcup\mathcal{A}^c$ satisfies the following area law
\begin{equation}\label{}
S_{\alpha}(\mathcal{A})\leq \zeta (K)\, \mathrm{Area}(\mathcal{A}),\,\, \forall \alpha,
\end{equation}
where $\mathrm{Area}(\mathcal{A})$ denotes the number of particles on the boundary of $\mathcal{A}$ and $\zeta(K)$ is a scaling factor only depends on the size of local cluster $K$.
\begin{pf}We first define three kinds of local clusters: (i) The clusters which only contain particles in $\mathcal{A}$ (as $\mathcal{C}_{int}$ in Fig. \ref{fig:entropy}), called internal clusters; (ii) The clusters which only contain particles in $\mathcal{A}^c$ (as $\mathcal{C}_{ext}$ in Fig. \ref{fig:entropy}), called external clusters; and (iii) The clusters which contain particles both in $\mathcal{A}$ and $\mathcal{A}^c$(as $\mathcal{C}_{bd}$ in Fig. \ref{fig:entropy}), called boundary clusters. We will denote the set of particles contained in boundary clusters as $\mathcal{B}=\partial \mathcal{A}\cup \partial \mathcal{A}^c$, where $\partial\mathcal{A}=\mathcal{A}\cap(\cup_{i}\mathcal{C}_{bd}^{(i)})$, $\mathcal{C}_{bd}^{(i)}$ are all boundary clusters, and similarly for $\partial\mathcal{A}^c$. The interior of $\mathcal{A}$, denoted as
$\mathrm{Int}\mathcal{A}$, is defined as $\mathrm{Int}\mathcal{A}=\mathcal{A}\backslash\partial \mathcal{A}$; the exterior of $\mathcal{A}$, denoted as $\mathrm{Ext}\mathcal{A}$, is defined as $\mathrm{Ext}\mathcal{A}=\mathcal{A}^c\backslash \partial \mathcal{A}^c$.

 Since
$\Psi(s_1,\cdots,s_N)=\prod_{i=1}^{M}\Phi_{i}(\mathcal{C}_i)$ are of quasi-product form, then using the locality feature of each local cluster, we have
\begin{align}\label{eq:pf}
|\Psi\rangle&=\sum_{s_1,\cdots,s_N}\prod_{i=1}^{M}\Phi_{i}(\mathcal{C}_i)|s_1\rangle\otimes \cdots\otimes|s_N\rangle\nonumber\\
&=\sum_{s_m\in \mathcal{B}}[\prod_{\mathcal{C}_{bd}^{(i)}\subset \mathcal{B}}\Phi_{i}(\mathcal{C}_{bd}^{(i)})]|\Phi_{L}(\partial \mathcal{A})\rangle\otimes|\Phi_R(\partial \mathcal{A}^c)\rangle
\end{align}
where $|\Phi_{L}(\partial \mathcal{A})\rangle=\sum_{s_n\in \mathrm{Int}\mathcal{A}}\prod_{\mathcal{C}_{int}^{(j)}\subset \mathcal{A}}\Phi(\mathcal{C}_{int}^{(j)})|\mathcal{A}\rangle$ and  $|\Phi_{R}(\partial \mathcal{A}^c)\rangle=\sum_{s_l\in \mathrm{Ext}\mathcal{A}}\prod_{\mathcal{C}_{ext}^{(j)}\subset \mathcal{A}^c}\Phi(\mathcal{C}_{ext}^{(k)})|\mathcal{A}^c\rangle$ ,  states $|\Phi_{L}(\partial \mathcal{A})\rangle$ are labeled by particles contained in $\partial\mathcal{A}$ and states $|\Phi_{R}(\partial \mathcal{A}^c)\rangle$ are labeled by the particles contained in $\partial \mathcal{A}^c$.
If each local system $s_i$ can take $p$ values,  there are at most  $p^{|\mathcal{B}|}$ terms contained in summation of Eq. (\ref{eq:pf}). We stress here that $|\mathcal{B}|$ only depends on the area  $\mathrm{Area}\mathcal{A}$  and cluster size $K$, more precisely, $|\mathcal{B}|\leq R\times\mathrm{Area}\mathcal{A}$. Therefore after tracing out the $\mathcal{A}^c$ part, we get $\rho_{\mathcal{A}}$ with rank at most $p^{|\mathcal{B}|}$, thus the R\'{e}nyi entropy $S_{\alpha}({\mathcal{A}})$ is upper bounded by $\zeta (K)\mathrm{Area}\mathcal{A}$.
\qed
\end{pf}
\end{thm}

For a physical system with a fixed background space-time, particles only interact with their neighborhoods, the strength of the interaction usually decays to zero if particles are sufficiently far apart. Thus for any many-body physical system, there is a preexisted geometry which characterizes the distance of two particles. However, when we write down a many-body quantum state, this part of geometrical information is usually erased in some sense. For a multipartite quantum system $\mathcal{S}$, we have a state $|\Psi_{\mathcal{S}}\rangle$ which encodes the full information of the system. To understand entanglement feature of $|\Psi_{\mathcal{S}}\rangle$, we first divide $\mathcal{S}$ into two parts $\mathcal{A}$ and $\mathcal{A}^c$, note that R\'{e}nyi entropy $S_{\alpha}(\mathcal{A})$ of the reduced density matrix $\rho_{\mathcal{A}}$ quantifies the entanglement between $\mathcal{A}$ and $\mathcal{A}^c$, then we can define entanglement feature of $|\Psi_{\mathcal{S}}\rangle$ as the set of R\'{e}nyi entropies $S_{\alpha}(\mathcal{A})$ over all entanglement subsystem $\mathcal{A}$. The entanglement feature of the system usually contains the information of the geometry of the system, namely, using the entanglement-geometry correspondence (duality), we can recover the geometry information from entanglement features.

Since the paradigm of the neural network is to adjust the connection weights, when weights are zero we say there is no connection between two neurons. It is then natural that the geometry of the neural network reflects in the connectivity of the network. In the spirit of entanglement-geometry correspondence, the entanglement is then encoded in the connectivity of neural networks of the state. This is consistent with the intuition we get from tensor network states, for which we can easily read out entanglement properties from the geometry of the tensor network. Since locally connected neural network states are all local quasi-product states, thus they all obey the entanglement area law. The locality of the states is encoded in the connection pattern of the neural network, which agrees well with our intuition. This kind of construction will also be useful for understanding the entanglement-geometry correspondence \cite{You2018}, such as Ryu-Takayanagi formula in a discrete form \cite{Ryu2006,Pastawski2015,almheiri2015bulk}.

Another issue worth mentioning is the topological entanglement entropy \cite{Kitaevtop}, which is a sub-leading term of $S(\mathcal{A})$. More precisely, for a gapped system, the entanglement entropy is expected to take the form $S(\mathcal{A})=\zeta \mathrm{Area} \mathcal{A}-\gamma+\mathcal{O}(|\mathrm{Area} \mathcal{A}|^{\beta})$, here $\zeta,\gamma,\beta\geq 0$, the first term is the area law contribution and $\gamma$ is called topological entanglement entropy. Topological entanglement entropy $S_{\mathrm{top}}=\gamma$ is a universal quantum number which can be used to detect if a system is topologically ordered or equivalently,  if the state is long-range entangled. The word \emph{universal} means that for the states with the same topological order, the topological entanglement entropy is the same. The local quasi-product state is a good ansatz state for gapped many-body system, thus it can also be used to calculate topological entanglement entropy. The procedure is as follows, for a local gapped quantum system $H=\sum_{k}H_k$, take the ansatz state as  $\Psi=\Pi_{k}\Phi_k(\mathcal{C}_k,\Omega_k)$, here $\mathcal{C}_k$ is the local cluster corresponding to $H_k$ which has the support (spins contained in $\mathcal{C}_k$) equal or lager than the support (spins which the operator acts nontivially) of $H_k$ and $\Omega_k$ is the variational parameters of the local cluster state $\Phi_k$ which, for example, can be chosen as the weights and biases for the $k$-th portion of local  neural network of the state. Using the variational method, we can find the value of parameters which minimize the energy functional and thus obtain the state. Then taking a large disc region $\mathcal{D}$ with smooth boundary, we can divide it as three fan-shaped subregions $\mathcal{A},\mathcal{B},\mathcal{C}$, the topological entanglement
 entropy can be obtained from
 $$S_{\text{top}}=S(\mathcal{A}\mathcal{B})+S(\mathcal{C}\mathcal{B})+S(\mathcal{A}\mathcal{C})-S(\mathcal{A})-S(\mathcal{B})-S(\mathcal{C})-S(\mathcal{D}).$$
As what we have shown in example \ref{example:toric} , the toric code state can be represented exactly as local quasi-product state in the most economic choice of local cluster, thus  $S_{\text{top}}$ of toric code model can be obtained exactly.
In general, to improve the accuracy of the calculation, the lager local cluster ought to be chosen.


\section{Entanglement features of deep neural network states}
\label{sec:DBM}
In this part, we investigate the entanglement properties of the deep neural network state, we will take the deep Boltzmann machine (DBM) as an example. Although many progress of RBM states has been made, the DBM states are less investigated \cite{gao2017efficient}. There are several crucial reasons why we need deep neural network rather than shallow one: (i) The representational power of shallow network is limited, there exist states which can be efficiently represented by deep neural network while the shallow one can not represent \cite{gao2017efficient}; (ii) Any Boltzmann machine (BM) can be reduced into a DBM, this also makes some limits in usage of shallow BM (with just one hidden layer, viz., RBM) \cite{gao2017efficient}; (iii) The hierarchy structure of deep neural is more suitable for encoding holography \cite{You2018,Gan2017,Hashimoto2019AdS} and for procedure such as renormalization \cite{Mehta2014}.

\begin{figure}
  \centering
  \includegraphics[width=9cm]{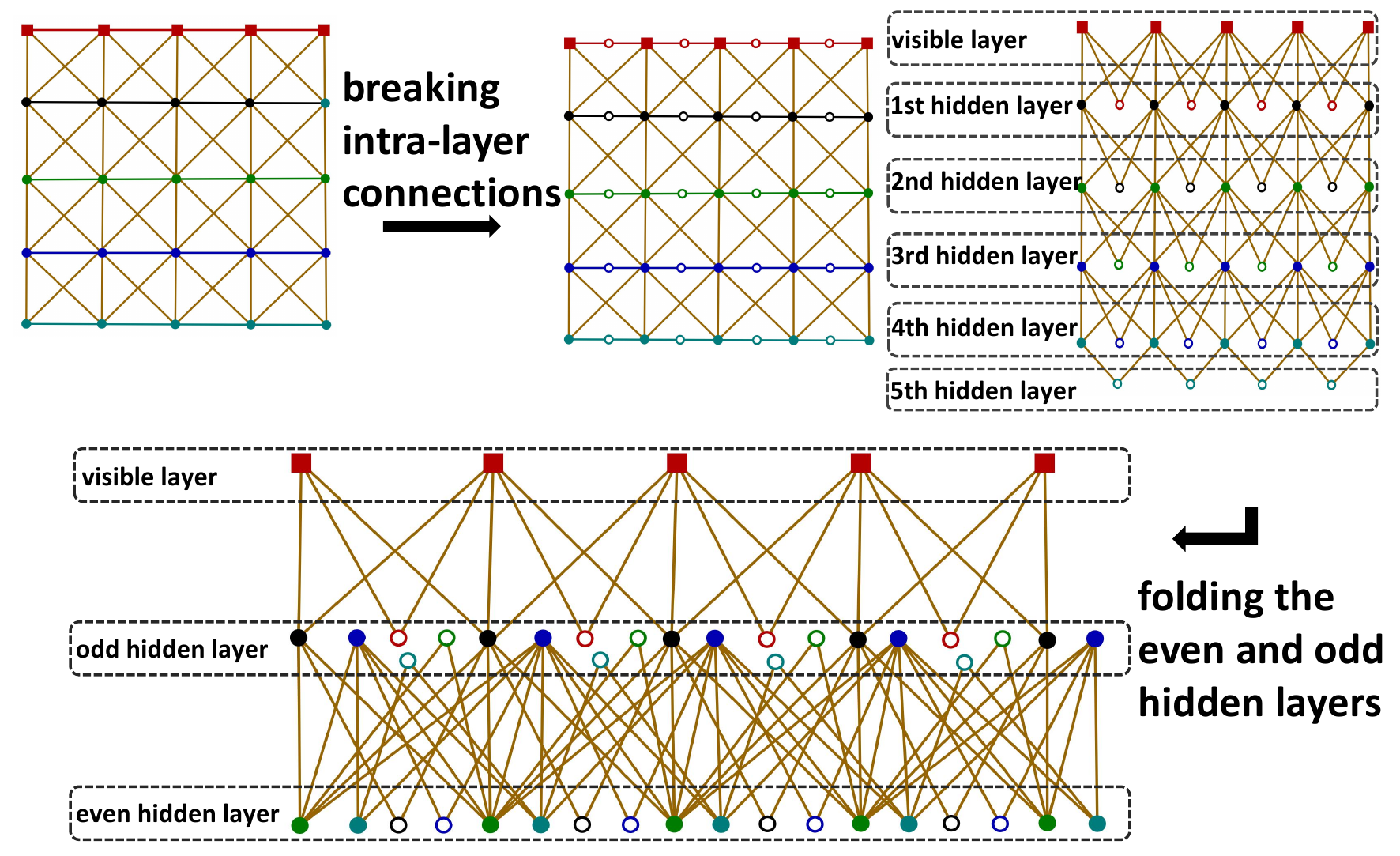}\\
  \caption{Illustration of the procedure for transforming an arbitrary Boltzmann machine into a deep Boltzmann machine with only two hidden layers. }\label{fig:folding}
\end{figure}

Now let us take a close look at the geometry of a DBM neural network. Since we can reduce a DBM with $M$ hidden layers into a DBM with only two hidden layers by folding trick \cite{gao2017efficient} (see Fig. \ref{fig:folding}), it is sufficient to consider the deep neural network with only two hidden layers. The procedure is the same as what has been done for shallow neural networks. The visible layer consists of physical variables (visible neurons), thus the geometry is given by the fixed background geometry (e.g., the lattice structure of the system). The geometry of the shallow and deep hidden layers are just duplicated from the visible lay geometry. Then we can define the distance between neurons not only in the same layer but also in different layers. For a given neuron $h$, the $\varepsilon$-neighborhood $B(h;\varepsilon)$ is defined as the the disk region centered at $h$ and with radius $\varepsilon$. An $\varepsilon$-local neural network is the one where each neuron only connects neurons in their $\varepsilon$-neighborhood, the maximum connecting number of each neuron is $K$, we call the network a local $K$-connecting (or $K$-local) DBM, see Fig. \ref{fig:layer} for an illustration.

\subsection{Entanglement area law}

Here we show that the geometrical information, \emph{locality}, of the deep neural network, also results in the area law of the entanglement entropy.
\begin{thm}
For any $K$-local DBM state $|\Psi\rangle$, R\'{e}nyi entropy of the reduced density matrix $\rho_{\mathcal{A}}=\mathrm{Tr}_{\mathcal{A}^c}(|\Psi\rangle\langle\Psi|)$ satisfy the following area law
\begin{equation}\label{}
S_{\alpha}(\mathcal{A})\leq \zeta(K)\, \mathrm{Area}(\mathcal{A}),\,\, \forall \alpha,
\end{equation}
where $\mathrm{Area}(\mathcal{A})$ denotes the number of particles on the boundary of $\mathcal{A}$ and $\zeta(K)$ is a scaling factor only depends on local connection number $K$.
\end{thm}
\begin{pf}
Inspired by the work of Deng \emph{et al.} \cite{Deng2017}, here we establish the area law by explicit construction. For K-local DBM states, we can group connections into several sets using the hidden neurons in the first hidden layer.  Note that for DBM states, the coefficients are $\Psi(\mathbf{v})= \sum_{\mathbf{h}}\sum_{\mathbf{g}}\exp\{\sum_i v_i a_i+\sum_k c_k g_k +\sum_j h_j(b_j+\sum_{i;\langle ij\rangle}W_{ij} v_i +\sum_{k;\langle kj\rangle}W_{kj} g_k)\}$ where sum is over hidden neurons $\mathbf{h}=(h_1,\cdots, h_m)$ of first hidden layer and $\mathbf{g}=(g_1,\cdots, g_l)$ of second hidden layer and all hidden neurons are assumed to take values $0,1$ here (of course, for the case of taking values $\pm 1$, the result also holds). Then, like what have usually been done for area-law tensor network states, we need to factorize the coefficients into a partial product form $\Psi(\mathbf{v})=\prod_{i=1}^ne^{v_ia_i}\sum_{\mathbf{g}}(\prod_{k=1}^le^{g_kc_k}\prod_{j=1}^m \Phi_j)$ in which $\Phi_j=\Phi_j(v_{j_1},\cdots,v_{j_K};g_{j_1},\cdots,g_{j_K})=\sum_{h_j=0,1}\exp\{ h_j(b_j+\sum_{i;\langle ij\rangle}W_{ij} v_i +\sum_{k;\langle kj\rangle}W_{kj} g_k)\}$, $v_{j_1},\cdots,v_{j_K}$ are (at most) $K$ visible neurons connected to $h_j$, and $g_{j_1},\cdots,g_{j_K}$ (at most)  $K$ hidden neurons connected to $h_j$ in deep hidden layer (where we have used the assumption of $K$-locality). Now using an important trick of Ref. \cite{Deng2017}, the visible layer neurons can be divided into six groups: $\mathcal{A}_3$ consists of visible neurons which connect $\mathcal{A}^c$ part by a hidden neuron, $\mathcal{A}_2$ consists of visible neurons connecting neurons of $\mathcal{A}_3$ via a hidden neuron, and $\mathcal{A}_1=\mathcal{A}\setminus(\mathcal{A}_3\bigcup \mathcal{A}_2)$; similarly, we can define $\mathcal{A}^c_3$, $\mathcal{A}^c_2$ and $\mathcal{A}^c_1$. Obviously, $\mathcal{A}$ ($\mathcal{A}^c$) is the disjoint union of $\mathcal{A}_1$, $\mathcal{A}_2$ and  $\mathcal{A}_1$ ($\mathcal{A}^c_3$, $\mathcal{A}^c_2$ and  $\mathcal{A}^c_1$). Similar division can be applied to the deep hidden layer (since the layer has a fixed background geometry same as visible layer), we first assume that the corresponding bipartition of the layer is $\mathcal{B}\sqcup \mathcal{B}^c$, the deep hidden neurons are then grouped into six parts: $\mathcal{B}_1$, $\mathcal{B}_2$, $\mathcal{B}_3$, $\mathcal{B}^c_3$, $\mathcal{B}^c_2$ and $\mathcal{B}^c_1$.

Now, consider the state $|\Psi\rangle=\sum_{\mathbf{v}}\Psi(\mathbf{v})
|\mathbf{v}_{\mathcal{A}}\rangle \otimes |\mathbf{v}_{\mathcal{A}^c}\rangle
=\sum_{\mathbf{v}}\prod_{i=1}^ne^{v_ia_i}\sum_{\mathbf{g}}(\prod_{k=1}^le^{g_kc_k}\prod_{j=1}^m \Phi_j)|\mathbf{v}_{\mathcal{A}}\rangle \otimes |\mathbf{v}_{\mathcal{A}^c}\rangle$, we denote the set of shallow hidden neurons which connect $\mathcal{A}_3$ and $\mathcal{A}^c_3$ (also $\mathcal{B}_3$ and $\mathcal{B}^c_3$) as $\mathcal{C}_{\mathrm{Bd}}$, ones connect $\mathcal{A}_1$ and $\mathcal{A}_2$ (also $\mathcal{B}_1$ and $\mathcal{B}_2$) as $\mathcal{C}_{\mathrm{Int}}$ and ones connect $\mathcal{A}^c_1$ and $\mathcal{A}^c_2$ (also $\mathcal{B}^c_1$ and $\mathcal{B}^c_2$) as $\mathcal{C}_{\mathrm{Ext}}$. We can introduce the state $|\Psi_{\mathcal{A}}\rangle=
\sum_{\mathbf{v}_{\mathcal{A}_1}}e^{\mathbf{v}_{\mathcal{A}_1}\cdot \mathbf{a}_{\mathcal{A}_1}}\sum_{\mathbf{g}_{\mathcal{B}_1}}e^{\mathbf{g}_{\mathcal{B}_1}\cdot \mathbf{c}_{\mathcal{B}_1}}\prod_{j\in \mathcal{C}_{\mathrm{Int}}}\Phi_j |\mathbf{v}_{\mathcal{A}}\rangle$ and state $|\Psi_{\mathcal{A}^c}\rangle=
\sum_{\mathbf{v}_{\mathcal{A}^c_1}}e^{\mathbf{v}_{\mathcal{A}^c_1}\cdot \mathbf{a}_{\mathcal{A}^c_1}}\sum_{\mathbf{g}_{\mathcal{B}^c_1}}e^{\mathbf{g}_{\mathcal{B}^c_1}\cdot \mathbf{c}_{\mathcal{B}^c_1}}\prod_{j\in \mathcal{C}_{\mathrm{Ext}}}\Phi_j |\mathbf{v}_{\mathcal{A}^c}\rangle$, then the state $\Psi$ can be decomposed as $$|\Psi\rangle=\sum_{\mathbf{v}_{\mathcal{A}_{\mathrm{Bd}}}}\sum_{\mathbf{g}_{\mathcal{B}_{\mathrm{Bd}}}}e^{\mathbf{v}_{\mathcal{A}_{\mathrm{Bd}}}\cdot \mathbf{a}_{\mathcal{A}_{\mathrm{Bd}}}}e^{\mathbf{g}_{\mathcal{B}_{\mathrm{Bd}}}\cdot \mathbf{c}_{\mathcal{B}_{\mathrm{Bd}}}}\prod_{j\in \mathcal{C}_{\mathrm{Bd}}}\Phi_j |\Psi_{\mathcal{A}}\rangle\otimes|\Psi_{\mathcal{A}^c}\rangle,$$
where $\mathcal{A}_{\mathrm{Bd}}=\mathcal{A}_2\cup\mathcal{A}_3\cup\mathcal{A}^c_3\cup\mathcal{A}^c_2$ and $\mathcal{B}_{\mathrm{Bd}}=\mathcal{B}_2\cup\mathcal{B}_3\cup\mathcal{B}^c_3\cup\mathcal{B}^c_2$.
Tracing out the $\mathcal{A}^c$ part, we get $\rho_{\mathcal{A}}$ is the weighted sum of several one-dimensional projectors (which are not necessarily orthogonal), via Gram-Schmidt orthogonalization, it is clear that the rank of $\rho_{\mathcal{A}}$ is upper bounded by a function $f(K)$ only depends on $K$. Since the R\'{e}nyi entropy of a density matrix take the maximum value only if all eigenvalues $p_i$ are equal, i.e., $p_1=\cdots=p_{r}=1/r\geq 1/f(K)$ with $r$ the rank of matrix, we then complete the proof. \qed
\end{pf}

Let us give some intuitive explanation about the construction. Since each visible neuron is correlated with neurons in $\varepsilon$-neighborhood, we can regard $\varepsilon$ as the correlation length of the state, the correlation between $\mathcal{A}$ and $\mathcal{A}^c$ comes predominantly from visible neurons sit in $2\varepsilon$-strip around the boundary $\partial\mathcal{A}$. Then the visible neurons deep in the region $\mathcal{A}$ can not be correlated with visible neurons deep in the region $\mathcal{A}^c$, only neurons near the boundary contribute to the R\'{e}nyi entanglement entropy $S_{\alpha}(\mathcal{A})$ and thus result in the area law.

\subsection{Entanglement volume law}
As we have proved above, the locality of the neural network reflects in the area law of entanglement features. It is natural to ask what about the neural network with nonlocal connections. We can expect the fully connected DBMs exhibit entanglement volume law \cite{Page1993,Foong1994,Sen1996}, actually this is the case. In contrast to tensor network for which the efficiency \footnote{We recall that the meaning of efficiency of a representation of many-body state is that the number of parameters to characterize the state increases at most polynomially at the number of particles $n$.} strongly depends on the validity of the entanglement area law of the state, the neural networks are still efficient in representing many-body states obeying volume law. As has been pointed out in Refs. \cite{Deng2017a,gao2017efficient}, shallow neural networks are capable of some critical-system states obeying entanglement volume law. We can trivially add a deep hidden layer and give some trivial connections to make a deep neural network exhibit volume law. Despite the triviality of the construction, we want to stress some crucial point of the volume-law neural network: (i) there must be some nonlocal connections in the neural network architecture, which is the origin of the volume law entanglement; (ii) the representation is efficient, i.e., the number of hidden neurons and connections increases at most polynomially at the number of visible neurons.

The volume-law DBM states have a close relationship with the maximally multipartite entangled states \cite{Facchi2008}. The philosophy behind the construction is that we can make the particles in the smaller region $\mathcal{A}$ fully correlate with its complementary $\mathcal{A}^c$ such that all information of $\mathcal{A}$ is encoded in $\mathcal{A}^c$ in some way, then $\rho_{\mathcal{A}}$ is proportional to identity matrix $\mathds{1}_{\mathrm{Vol}(\mathcal{A})}$ with order $\mathrm{Vol}(\mathcal{A})$ (where $\mathrm{Vol}(\mathcal{A})$ denotes the number of particles contained in region $\mathcal{A}$), which further implies that R\'{e}nyi entropies satisfy the volume law. Another important issue is that the number of the neural network is closely related to the entanglement properties of the corresponding neural network states. A neural network state with more hidden layers will tend to exhibit the volume law entanglement. This can be seen from feed-forward neural network more easily, if the number of the hidden layers increases, eventually, the size of the local clusters will be comparable with the system size, this will break the entanglement area law.

\section{Understanding the power of neural network using R\'{e}nyi entanglement entropy}
\label{sec:image}
The success of neural network achieved in tasks like image classification problem suggests that to understand the power of the neural network we will need to establish a new information theory of the functions of images rather than just of the image itself.
In fact the classification problem shares great mathematical similarity with the quantum spin model \cite{zhang2017entanglement}.
The images correspond to the spin configurations and the target function of the image classification problem corresponds to the wave function of the spin system.  The functions $f$ of the images with $N=L\times L$ pixels form a Hilbert space, and by normalization, these functions are in one-to-one correspondence with the wave functions of a $L\times L$ quantum spin model, thus the notion of R\'{e}nyi entanglement entropy $S_{\alpha}(\mathcal{A})$ of a connected subregion of image makes sense.
As we will see, this entanglement entropy can be used as a measure of the difficulty for approximating a function. In general, functions obey the entanglement volume law need $O(2^N)$ parameters to approximate, while functions obey entanglement area law can be approximated using a neural network with $\mathsf{poly}(N)$ parameters. Here we will argue that entanglement entropy of target functions of reasonable image classification problems obey the entanglement area law.

We known from quantum spin model that to represent a general quantum state using, e.g.,  tensor network and neural network, $O(2^N)$ parameters are needed \cite{Page1993,Foong1994,Sen1996}, but with the locality constraint of the Hamiltonian, polynomially many parameters are sufficient  to represent the ground state, this is characterized by the area law of entanglement entropy of these states \cite{Eisert2010}.

Following the Ref. \cite{zhang2017entanglement}, here we first present the explicit definition of image classification problem. The case we will consider here a is two-label classification problem (a \emph{yes} or \emph{no} problem) of the $L\times L$-pixel black-white images, the corresponding values of pixels are thus $1$ and $0$ for black and white respectively. The set of all images is denoted as $\mathcal{I}=\{I:\{1,\cdots,L\}\times \{1,\cdots,L\}\to\{0,1\}\}$, there are $2^{L\times L}$ images in $\mathcal{I}$ in total. The Hilbert space of all complex-valued functions on image set $\mathcal{I}$ is then $\mathcal{H}_{\mathcal{I}}=\{f:\mathcal{I}\to \mathbb{C}\}$. The classification problem is to determine if a given image $I\in\mathcal{I}$ lies in the target set $\mathcal{T}\subset \mathcal{I}$ or not. The target function is then defined as
\begin{equation}\label{eq:target}
f_{\mathcal{T}}(I)=\left\{
\begin{array}{ll}{1,} & {\text { if } I\in \mathcal{T}}, \\
{0,} & {\text { otherwise }}.
\end{array}\right.
\end{equation}
To measure the difference between general $f$ and target function $f_{\mathcal{T}}$, some kind of norm $||f(I)-f_{\mathcal{T}}(I)||$ is chosen, and we can construct a functional called cost function
$C_{\mathcal{T}}[f]=\sum_{I\in \mathcal{I}}||f(I)-f_{\mathcal{T}}(I)||p(I)$ where $p(I)$ is the probability distribution over the image set. Our aim is to minimize the cost function, i.e.,
$$\textbf{Image classification problem:}\,\, \mathrm{argmin}\{f\in \mathcal{H}_{\mathcal{I}}|C_{\mathcal{T}}[f]\}.$$

It is extremely difficult to solve the optimization problem directly. In the neural network approach, function $f$ is represented by a neural network (with fixed architecture) with parameter set $\Omega=\{w_{ij},b_i\}$, i.e., $f(I)=f(\Omega,I)$, the cost function then becomes a function of these parameters $C_{\mathcal{T}}[f]=C_{\mathcal{T}}(w_{ij},b_i)$, algorithms like gradient descent method can then be applied to find the minimal value.

Let us see an example of the image classification problem for which an exponentially large number of parameters are needed to approximate the corresponding target function. Suppose that we have randomly generated a set of $L\times L$-pixel images and set them as the target set $\mathcal{T}$ of images, then $\mathcal{T}$ does not have any pattern at all. It turns out that this problem can only be solved with exponentially many parameters and the target function obeys the volume law of entanglement \cite{Page1993,Foong1994,Sen1996}. By contrast, problems whose target set has an intrinsic pattern can be solved using polynomially many parameters. For example, determining whether a given image only contains loops or not can be solved efficiently and the target function obeys the entanglement area law (which in fact corresponds to the toric code model in quantum spin models \cite{Kitaev2003}). It is natural to ask what kind of conditions the target set of images must satisfy to make the classification problem can be solved efficiently. In Ref. \cite{zhang2017entanglement}, two important conditions are given: (i) For images $I$ and $I'$, consider two connected regions $\mathcal{A}$ and $\mathcal{A}^c$, if $I$ and $I'$ are the same in region in $\mathcal{A}^c$ then they must be the same on the boundary of $\mathcal{A}$; (ii) For two regions $\mathcal{A}_{I}$ and $\mathcal{A}_{I'}$ of images $I$ and $I'$, the number $N_{I,I'}$ of possible images, for which $\mathcal{A}^c_{I}$ and $\mathcal{A}^c_{I'}$ are the same, only depends on the $B$-range part of boundary of  $\mathcal{A}_{I}$ and $\mathcal{A}_{I'}$. These two conditions characterize the property of smoothness of the images in some sense, we will refer to this kind of problem as a locally smooth image classification problem. From the above discussion, we may come to the following result:
\begin{thm}
For any target function $f_{\mathcal{T}}$ of locally smooth image classification problem, the R\'{e}nyi entanglement entropy satisfies the area law
\begin{equation}
S_{\alpha}(\mathcal{A})\leq \zeta(B) \mathrm{Area}(\mathcal{A}),
\end{equation}
where $\zeta(B)$ is a scaling factor depending on $B$ linearly and not depending on the size of the images (number of pixels).
\end{thm}
\begin{pf}
This can be proved straightforward by calculating the density matrix corresponding to target function $f_{\mathcal{T}}$, and by tracing the region $\mathcal{A}^c$ part, we obtain a density matrix $\rho_{\mathcal{T}}(\mathcal{A})$ with rank at most $2^{\zeta(B)\times \mathrm{Area}(\mathcal{A})}$.  To this end, we first introduce the density matrix $\rho_{\mathcal{T}}$ corresponding to target function $f_{\mathcal{T}}$. We can assign $|0\rangle$ or $|1\rangle$ to each pixel $I_{ij}$ of the image $I$ when it is white $I_{ij}=0$ or black $I_{ij}=1$ respectively, in this way, we construct a $2^{N}$-dimensional Hilbert space with basis $|I\rangle=\otimes_{i,j=1}^{L}|I_{ij}\rangle$. Then for a target function $f_{\mathcal{T}}$ as in Eq. (\ref{eq:target}) we can construct a quantum state 
$$|\psi_{\mathcal{T}}\rangle=\frac{1}{\mathcal{N}}\sum_{I\in \mathcal{I}}f_{\mathcal{T}}(I)|I\rangle=\frac{1}{\mathcal{N}}\sum_{I\in\mathcal{T}}|I\rangle,$$
where $\mathcal{N}$ is the normalization factor.

Now consider a connected region $\mathcal{A}$ of the images, let $I_{\mathcal{A}}$ and $I_{\mathcal{A}^c}$ denote pixels in region $\mathcal{A}$ and $\mathcal{A}^c$ respectively. Thus $|\psi_{\mathcal{T}}\rangle=\sum_{I\in\mathcal{T}}|I\rangle/\mathcal{N}=\sum_{I\in\mathcal{T}}|I_{\mathcal{A}}\rangle\otimes|I_{\mathcal{A}^c}\rangle/\mathcal{N}$.
To ultilize two conditions of the target images of locally smooth image classification problem, we further divide the region $\mathcal{A}$ into $\partial^B \mathcal{A}$ which contains pixels in $B$-rangle of the boundary of $\mathcal{A}$, $\mathrm{Int}^B(\mathcal{A})=\mathcal{A}\setminus \partial^B \mathcal{A}$. The target state can be rewrite as $|\psi_{\mathcal{T}}\rangle=\sum_{I\in\mathcal{T}}|I_{\partial^B \mathcal{A}}\rangle\otimes |I_{\mathrm{Int}^B(\mathcal{A})}\rangle\otimes |I_{\mathcal{A}^c}\rangle/\mathcal{N}$. The reduced density matrix is
\begin{align}\label{eq:targetstate}
\rho_{\mathcal{T}}(\mathcal{A})&=\mathrm{Tr}_{\mathcal{A}^c}|\psi_{\mathcal{T}}\rangle\langle \psi_{\mathcal{T}}|,\nonumber\\
&=\frac{1}{\mathcal{N}^2}\sum_{I_{\mathcal{A}},I'_{\mathcal{A}};I,I'\in\mathcal{T}} 
\sum_{I_{\mathcal{A}^c},I'_{\mathcal{A}^c};I,I'\in\mathcal{T}}
|I_{\mathcal{A}}\rangle \langle I'_{\mathcal{A}}| \delta_{I_{\mathcal{A}^c},I'_{\mathcal{A}^c}}.\nonumber
\end{align}
Now we can divide the target image set $\mathcal{T}$ into $K$ disjoint subsets $\mathcal{T}_j,j=1,\cdots,K$ such that for any two images $I,I'\in\mathcal{T}_j$, they share the same pixel values in regions $\mathcal{A}^c$, i.e., $I_{\mathcal{A}^c}=I'_{\mathcal{A}^c}$. Thus the reduced density matrix takes the form (the normalization factor is omitted here)
\begin{align}
&\sum_{j=1}^{K}\sum_{I_{\mathcal{A}},I'_{\mathcal{A}};I,I'\in\mathcal{T}_j}|I_{\mathcal{A}}\rangle \langle I'_{\mathcal{A}}|, \nonumber\\
=&\sum_{j=1}^{K}\sum_{I_{\mathcal{A}},I'_{\mathcal{A}};I,I'\in\mathcal{T}_j}| I_{\mathrm{Int}^B(\mathcal{A})}\rangle |I_{\partial^B\mathcal{A}}\rangle\langle I'_{\mathrm{Int}^B(\mathcal{A})}|\langle I'_{\partial^B\mathcal{A}}|.\nonumber
\end{align}
Then we need to use conditions (i) and (ii) of  locally smooth image classification problem, from condition (i), we know that $K$ is upper bounded by $2^{\mathrm{Area}(\mathcal{A})}$, since images with the same $\mathcal{A}^c$ values must also share the same $\partial \mathcal{A}$ value, the number of possible $\mathcal{A}^c$ images in target set $\mathcal{T}$ is upper bounded by the possible images of boundary of $\mathcal{A}$, viz., $K\leq 2^{\mathrm{Area}(\mathcal{A})}$. From condition (ii), for a chosen $\mathcal{T}_j$, the number of possible $\mathrm{Int}^B (\mathcal{A})$ values for images in $\mathcal{T}_j$ is upper bounded by the number of possible images of the $B$-range boundary of $\mathcal{A}$, thus there are at most $2^{B\mathrm{Area}(\mathcal{A})}$ terms in the summation
\begin{align}
&\sum_{I_{\mathcal{A}},I'_{\mathcal{A}};I,I'\in\mathcal{T}_j}| I_{\mathrm{Int}^B(\mathcal{A})}\rangle |I_{\partial^B\mathcal{A}}\rangle\langle I'_{\mathrm{Int}^B(\mathcal{A})}|\langle I'_{\partial^B\mathcal{A}}|\nonumber \\
=&\sum_{I_{\mathrm{Int}^B(\mathcal{A})},I'_{\mathrm{Int}^B(\mathcal{A})};I,I'\in\mathcal{T}_j}| I_{\mathrm{Int}^B(\mathcal{A})}\rangle |I_{\partial^B\mathcal{A}}\rangle\langle I'_{\mathrm{Int}^B(\mathcal{A})}|\langle I_{\partial^B\mathcal{A}}|.\nonumber
\end{align}
Thus, it is clear that the rank of density matrix $\rho_{\mathcal{T}}(\mathcal{A})$ is upper bounded by $2^{(B+1)\mathrm{Area}(\mathcal{A})}$, the area law of R\'{e}nyi entropy is now established and $\zeta(B)=B+1$.
\qed
\end{pf}
It is worth mentioning that the condition (ii) about the $B$-range boundary determines area law and $B$ must not depend on the number of pixels of the images.
Let us now see a typical example of the locally smooth image classification problem.
\begin{example}[Circle determination on a torus $\mathbb{T}^2$]
Consider a set of $L\times L$-pixel images, our aim is to determine if the image is a circle or not. Here we take the periodic boundary condition for images, thus each image can be regarded as a torus image. To make the definition of the circle more clear, let us consider the cellulation of a torus $\mathbb{T}^2$, which is  chosen as a square lattice on the torus, thus there are three kinds of cells here, faces $f\in F(\mathbb{T}^2)$, edges $e\in E(\mathbb{T}^2)$ and vertices $v\in V(\mathbb{T}^2)$. The pixels are put on the edges of the lattice, viz., we assign $0,1$ to each edge  (See Fig. \ref{fig:GSDloop}).  

In homology-theoretical language, a torus image is actually a $\mathbb{Z}_2$ one-chain, which is defined as $I: E(\mathbb{T}^2)\to \{0,1\}$. Likewise, we can defined the zero-chain $c: V(\mathbb{T}^2)\to \{0,1\}$, a trivial zero-chain is the one for which all vertices are set as $0$. The set of all one-chain and zero-chain is denoted as $C^{1}(\mathbb{T}^2)$ and $C^{0}(\mathbb{T}^2)$ respectively.
Consider a one-chain, we can define a boundary map $\partial :C^{1}(\mathbb{T}^2) \to C^{0}(\mathbb{T}^2)$ which maps the boundary value of the edge to the vertices connected by the edge, and for the vertex $v$ connecting edges $e$ and $e'$, the value of $v$ is defined as the the addition of values of $e$ and $e'$ (this addition is modulo two). For example, consider an image (one-chain) $I$ for which $I(e)=1$ for some $e$ and all other edges are set $0$. If the edge $e$ has two end vertices $u,v$, then the zero-chain $\partial I$ is the one for which $v,u$ take value $1$ and all other vertices take value $0$. Now we are at a position to define the \emph{circle}, a circle (or more rigorously a one-circle)  is defined as a one chain $I$ for which $\partial I$ is a trivial zero-chain. The set of all circle images is denoted as $Z^1(\mathbb{T}^2)$ which is nothing but the target set of the circle determination problem.
The state corresponding the target function of the problem is defined as
\begin{equation}
|\psi\rangle =\frac{1}{\mathcal{N}}\sum_{I\in Z^1(\mathbb{T}^2)} |I\rangle,
\end{equation}
which is a local quasi-product state, thus obeying the entanglement area law, see example \ref{example:toric}.
\end{example}

\section{Conclusion and discussion}

In this paper, the authors establish the entanglement area law of the shallow and deep neural network states. By introducing the notion of locality into the neural network representations of quantum states, we can see that the resulting local neural network states obey the entanglement area law. It is worth mentioning that there are some subtle issues in the construction here.

The first crucial issue we want to discuss is the topology of the neural network states. Each $L$-layer neural network architecture corresponds to a $L$-partite graph $G$, in a $L$-partite graph, the vertex set $E(G)$ are divided into $L$ disjoint subsets, which correspond to the different layers of the neural network, and there is no edges in each of these subsets (no intra-layer connections in neural network language), and we say that two neural networks are equivalent if the corresponding graphs are isomorphic, we denote the isomorphic class of neural network as $\mathcal{G}$. The neural network in an isomorphic class have the same representational power and the corresponding quantum states have completely the same physical properties. In our construction, we fix the geometry of the physical layer of the neural network, and use this fixed background geometry, we introduce the notion of locality into the neural network, since every neural network which is equivalent to this network with the fixed background geometry have the same topology, thus the properties of the state only depends on the topology not geometry of the state. This means that each time we want to study the neural network states represented by neural networks given by the equivalent class $\mathcal{G}$, we  can choose a representative from $\mathcal{G}$ without loss of generalities, and for convenience, we can always choose the one with fixed background geometry.

Another issue we want to stress is the neural network approach to AdS/CFT correspondence, or more precisely, entanglement-geometry correspondence in this context. To realize the Ryu-Takayanagi formula, we first construct a neural network whose physical layer corresponds to the boundary physical degrees of freedoms, and the bulk geometry is given by the hidden layers and connections between neurons. The essential idea behind this approach to holography (or entanglement-geometry correspondence) is that the entanglement feature is encoded in the neural network geometry. That is, the bulk geometry is given, and the neural network is tiled on the background bulk geometry, the structure of the neural network provides the required entanglement features on the boundary which is exactly the dual of the bulk geometry. Note that in Ref. \cite{You2018}, the inverse problem of the above problem is investigated, where they explored how to use given entanglement features of a state to determine the optimal holographic geometry. These topics will be left for our future studies. During the preparation of our manuscript, we notice that a related work was made available \cite{Hashimoto2019AdS}.

\begin{acknowledgments}
Z. A. Jia and L. Wei are of equal contribution to this work.
We acknowledge Dong-Ling Deng for discussions during his visiting at USTC, we also acknowledge I. Glasser for bringing our attentions to the works \cite{Glasser2018a,glasser2018supervised,robeva2018duality}, where  special case of local quasi-product state is given (the local cluster states are of MPS form or more general tensor network form). Z. A. Jia acknowledges Zhenghan Wang and the math department of UCSB for hospitality.
This work was supported by the National Key Research and Development Program
of China (Grant No. 2016YFA0301700), and the Anhui Initiative in Quantum Information Technologies
(Grants No. AHY080000)
\end{acknowledgments}


\bibliographystyle{apsrev4-1-title}

%

\end{document}